\documentclass[twocolumn,amsmath,amssymb]{revtex4}

\def\gsim{ \lower .75ex \hbox{$\sim$} \llap{\raise .27ex \hbox{$>$}} }

\def\lsim{ \lower .75ex \hbox{$\sim$} \llap{\raise .27ex \hbox{$<$}} }

\usepackage{graphicx}
\usepackage{dcolumn}
\usepackage{bm}
\input epsf
\def\gsim{ \lower .75ex \hbox{$\sim$} \llap{\raise .27ex \hbox{$>$}} }
\def\lsim{ \lower .75ex \hbox{$\sim$} \llap{\raise .27ex \hbox{$<$}} }

\usepackage{graphicx}
\usepackage{dcolumn}
\usepackage{bm}

\newcommand{\nn}{\nonumber}
\newcommand{\be}{\begin{equation}}
\newcommand{\ee}{\end{equation}}
\newcommand{\bea}{\begin{eqnarray}}
\newcommand{\eea}{\end{eqnarray}}

\def\d{\delta}

\def\s{\sigma}
\def\e{\epsilon}

\def\f{\phi}

\def\k{\kappa}

\def\s{\sigma}

\def\t{\tau}

\begin{document}

\title{Non-Gaussian Density Fluctuations from Entropically Generated Curvature
Perturbations in Ekpyrotic Models}

\author{Jean-Luc Lehners$^{1}$ and Paul J. Steinhardt$^{1,2}$}
\affiliation{ $^1$Princeton Center for Theoretical Physics, Princeton
University, Princeton, NJ 08544 USA \\
$^2$Joseph Henry Laboratories, Princeton
University, Princeton, NJ 08544,
USA
}

\begin{abstract}
We analyze the non-gaussian density perturbations generated in
ekpyrotic/cyclic models based on heterotic M-theory. In this
picture, two scalar fields produce nearly scale-invariant
entropic perturbations during an ekpyrotic phase that are
converted into curvature modes {\it after the ekpyrotic phase
is complete} and just before the big bang. Both
intrinsic non-linearity in the entropy perturbation and the
conversion process contribute to non-gaussianity.
The range of
the non-gaussianity parameter $f_{NL}$ depends on the details of the scalar field potential during the ekpyrotic phase, and on
how gradual the conversion process is.   Although a wider range is
possible, in principle,
natural values of the parameters of the potential combined
with a gradual conversion process lead to values of $-60 \lesssim f_{NL} \lesssim +80$, typically much greater than slow-roll inflation but
within the current observational bounds.

\end{abstract}


\maketitle

Ekpyrotic \cite{Khoury:2001wf} and cyclic
\cite{Steinhardt:2001vw} models of the universe use quantum
fluctuations of scalar fields produced during a slowly
contracting phase with equation of state $w>1$ to generate the
observed nearly scale-invariant spectrum of curvature (energy
density) fluctuations after the big bang. One mechanism
considered for converting the fluctuations of a scalar field
into cosmological curvature perturbations relies on
higher-dimensional effects at the collision between orbifold
planes (branes) along an extra dimension \cite{Tolley:2003nx}.
Recently, however, a new ``entropic" mechanism
\cite{Lehners:2007ac,others} has been proposed that relies on
two (or more) scalar fields and ordinary 4d physics,
stimulating new approaches to ekpyrotic and cyclic cosmology
that may not require branes or extra dimensions at all
\cite{Buchbinder:2007ad,Creminelli:2007aq,Koyama:2007mg,Battefeld:2007st}.

In this paper, we wish to consider an important byproduct of
the entropic mechanism: a non-gaussian contribution to the
density fluctuation spectrum that is more than an order of
magnitude greater than for conventional inflationary models and
that can satisfy current observational bounds, including the
recently claimed detection of non-gaussianity
\cite{Yadav:2007yy}. Our results differ significantly from the
cases considered by Koyama {\it et al.}\cite{Koyama:2007if} and
Buchbinder {\it et al.}\cite{Buchbinder:2007at} in which they
assumed an ekpyrotic ($w \gg 1$) phase that continues all the
way to the conversion of entropic to curvature fluctuations, as
in the ``new ekpyrotic" model \cite{Buchbinder:2007ad}; for
these cases, the non-gaussianity is amplified by the non-linear
evolution on super-horizon scales to the point where $f_{NL}$,
the parameter characterizing the non-linear curvature
perturbation \cite{Komatsu:2000vy}, reaches magnitude ${\cal O}
(c_1^2)$, where $c_1 \approx 2\sqrt{w+1}$ parameterizes the
steepness of the scalar field potential during the ekpyrotic
phase. A potential problem is that a minimum of $c_1 \ge 10$ is
required just to satisfy the current upper bound constraints on
$n_s$, the spectral tilt of the scalar (energy density)
perturbation spectrum \cite{Lehners:2007ac}; and $c_1 \ge 30$
is needed to reach the best-fit value $n_s \approx 0.97$. Yet,
excluding finely-tuned cancellations, the resulting value of
$f_{NL}$ obtained in
Ref.~\cite{Buchbinder:2007ad,Creminelli:2007aq,Koyama:2007mg,Battefeld:2007st}
is marginally consistent with current observational bounds on
non-gaussianity only for $c_1$ restricted to a narrow range
($\lesssim 15$).

Here we show that, by having the conversion occur after the
ekpyrotic phase, as is natural from the point of view of
heterotic M-theory and the cyclic model
\cite{Steinhardt:2001vw,Lehners:2007ac}, the non-gaussianity is
reduced and much less sensitive to $c_1$, so that one can
naturally fit the best-fit value of $n_s$ and current bounds on
both $f_{NL}$ at the same time. More specifically, the net
value of $f_{NL}$ depends on both the intrinsic non-linearity
in the entropy perturbation as it is generated during the
ekpyrotic phase as well as the duration of the conversion
process. These two contributions to $f_{NL}$ compete with each
other, in that the intrinsic non-linearity, which depends on
the steepness of the scalar field potential, is always positive
and the contribution depending on the duration of the
conversion process is negative. In principle, with arbitrarily
steep potentials and arbitrarily sharp conversions, a wide
range of $f_{NL}$ is possible, and one can even have a
finely-tuned cancellations of two large but oppositely signed
contributions that give nearly zero non-gaussianity. Excluding
these fine-tuned cases, though, we find that
 $f_{NL}$ lies within the current
observationally favored range
 of
 $26.9 < f_{NL} < 146.7$
that was recently reported at $2\s$ \cite{Yadav:2007yy} or the bound
obtained earlier by the WMAP collaboration, $-36  < f_{NL} < 100$
\cite{Spergel:2006hy}, while, at the same time, matching the
best-fit value of $n_s$
 allowed by WMAP.

As a concrete example, we will analyze the 4d effective field
theory derived from the ekpyrotic/cyclic model of
\cite{Lehners:2007ac}. The 4d theory describes the phase when
two boundary branes approach one another and collide
\cite{Lehners:2006pu} in heterotic M-theory \cite{LOSW1}; for a
short review see \cite{Lehners:2007gy}. The brane worldvolume
is described by gravity and two canonically normalized scalar
fields, $\phi_{1,2}$, which parameterize the distance between
the two branes as well as the volume of the internal Calabi-Yau
space \cite{Lehners:2006pu}. For the purposes of this paper,
though, the connection to M theory is not essential; it simply
provides physical motivation for considering this particular 4d
field theoretic example. The same effects will occur in other
4d field theories in which the conversion of entropic to
curvature perturbations occurs after the ekpyrotic ($w \gg 1$)
phase.

We assume a flat Friedmann-Robertson-Walker (FRW) background
with line element $\d s^2= -\d t^2 +a^2(t) \d {\bf x}^2$ and
scale factor $a(t)$, and steep, negative, scalar field
potentials of the form \be V(\phi_1, \phi_2)= -V_1 e^{-\int c_1
\d \phi_1}-V_2 e^{-\int c_2 \d\phi_2},
\label{potentialoriginal}\ee where $c_1=c_1(\phi_1)$, $c_2 =
c_2(\phi_2)$ and $V_1$,$V_2$ are positive constants. During the
ekpyrotic phase, the $c_i$ are nearly constant and  $c_i \gg
1$.  Then the Einstein-scalar equations admit a scaling
solution \be a = (-t)^{1/\e}, \quad \phi_i = {2\over c_i} \ln
(-\sqrt{c_i^2 V_i/2} \, t),
 \quad \frac{1}{\e}= \sum_i {2 \over c_i^2},
\label{c2} \ee which describes a very slowly contracting
universe with $\e \gg 1$ and $w+1 = 2\e/3 \gg1$ -- the defining
characteristics of an ekpyrotic phase. At the end of the
ekpyrotic phase, the effective values of $c_i \rightarrow 0$
and the ekpyrotic potential energy ceases to be important
cosmologically.

\begin{figure}[t]
\begin{center}
\includegraphics[width=0.45\textwidth]{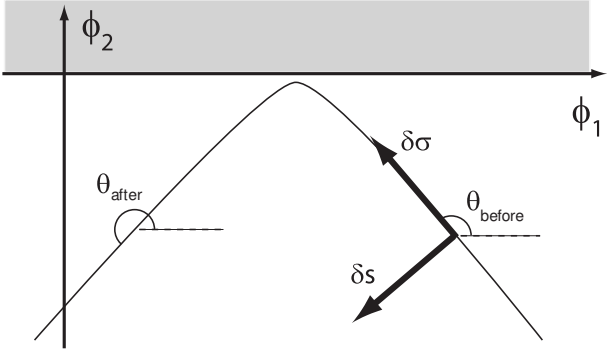}
\caption{\label{Figure1} {\small
The background trajectory in scalar field space is
straight everywhere except for a brief period after the ekpyrotic phase
around the time of reflection. The shaded region indicates the geometrically
forbidden region bounded by the axis $\phi_2=0.$ In the presence of brane-bound
matter, the trajectory reflects smoothly off this boundary. Also shown are the
directions of adiabatic ($\delta \sigma$) and entropic ($\delta s$)
fluctuations, as well as the angle of the trajectory before and after the
reflection.}}
\end{center}
\end{figure}

The colliding branes solution corresponds to an almost
everywhere straight line trajectory in scalar field space in
the 4d effective theory. The only deviation from this straight
trajectory is one required by the fact that the scalar field
moduli space admits a boundary at $\phi_2=0,$ which from the
higher-dimensional point of view corresponds to the locus where
both the scale factor on the negative-tension brane and the
volume of the Calabi-Yau manifold reach zero. As shown in
\cite{Lehners:2006pu}, the trajectory simply gets reflected at
this boundary. In the presence of matter on the branes, the
reflection is smoothed out by an effective repulsive potential
\cite{Lehners:2006pu} and the trajectory never actually touches
the boundary, as shown in Fig.~\ref{Figure1}. This smooth
reflection is relevant for the density fluctuation spectrum,
because it suffices to induce the conversion from entropic to
curvature perturbations without introducing any extra elements
to the theory. In fact, one could even imagine that it is the
effective potential associated with this reflection which
causes the ekpyrotic phase to end. In any case, the conversion
is followed by a period of scalar field kinetic energy
domination, which sets simple cosmological conditions near the
big crunch/big bang transition \cite{Steinhardt:2001vw}.

It is useful to recast the evolution in terms of the adiabatic
variable $\s$ (pointing along the background trajectory) and
the entropy variable $s$, pointing transverse to the trajectory
(see Fig.~\ref{Figure1}) \cite{Langlois:2006vv}. Up to
unimportant additive constants that will be fixed shortly, we
define \bea \s &\equiv& \frac{\dot\phi_1 \phi_1 + \dot\phi_2
\phi_2}{\dot\s}
\\ s &\equiv& \frac{\dot\phi_1 \phi_2 - \dot\phi_2
\phi_1}{\dot\s}, \eea where $\dot{\s} \equiv
\sqrt{\dot{\phi}_1^2+\dot{\phi}_2^2}$ ($\dot{\s}$ is positive
by definition). A dot represents a derivative with respect to
FRW time. Then we can expand the potential up to third order as follows
\cite{Buchbinder:2007tw}: \be V_{ek}=-V_0 e^{\sqrt{2\e}\s}[1+\e
s^2+\frac{\k_3}{3!}\e^{3/2} s^3],
\label{potentialParameterized}\ee where $\k_3$ is of ${\cal O}(1)$
 for typical potentials (the case of exact exponentials corresponds to $\k_3=-4\sqrt{2/3}$).
The scaling
solution (\ref{c2}) can be rewritten as \be a(t)=(-t)^{1/\e} \qquad
\s=-\sqrt{\frac{2}{\e}}\ln \left(-\sqrt{\e V_0} t\right) \qquad
s=0. \label{ScalingSolution}\ee The perturbations about this
contracting universe have a growing mode -- the entropy
perturbation --  corresponding to the relative fluctuation of
the two fields:  $\delta s \equiv (\dot{\phi}_1\,\delta \phi_2
- \dot{\phi}_2\,\delta \phi_1)/\dot{\s},$  We will decompose
the entropy perturbation into a linear, gaussian part and a
second-order perturbation by writing $\delta s = \delta s^{(1)}
+ \delta s^{(2)}.$ Its equation of motion, on large scales and
up to second order in field perturbations is then given by
\cite{Langlois:2006vv} \bea && \ddot{\delta s} +3H \dot{\delta
s}+
\left(V_{ss}+3\dot{\theta}^2 \right) \delta s \nn \\
&&+\frac{\dot{\theta}}{\dot{\sigma}}(\dot{\delta s}^{(1)})^2
+\frac{2}{\dot{\sigma}}\left( \ddot{\theta}+
\dot{\theta}\frac{V_{\sigma}}{\dot{\sigma}} -
\frac{3}{2}H\dot{\theta}\right)\delta s^{(1)} \dot{\delta
s}^{(1)} \nn \\ && +\left( \frac{1}{2}
V_{sss}-\frac{5\dot{\theta}}{\dot{\sigma}}V_{ss}-
\frac{9\dot{\theta}^3}{\dot{\sigma}} \right)(\delta s^{(1)})^2
+\frac{2\dot{\theta}}{\dot{\sigma}}\delta \epsilon^{(2)}= 0.
\label{eq-entropy} \eea Here $H \equiv \dot{a}/a$ is the Hubble parameter,
$V_{\sigma}$ denotes a derivative
of the potential along the background trajectory and
$V_{s\cdots s}$ denote successive derivatives along the $\delta
s$ direction. The angle $\theta$ of
the background trajectory is defined as in \cite{Gordon:2000hv}
by $\cos(\theta)=\dot{\f}_1/\dot{\s}, \, \sin(\theta) =
\dot{\f}_2/\dot{\s}.$ A useful expression for the time
variation of this angle is $\dot{\theta} = -V_{s}/\dot{\s}.$ In
the simplest models, the background scalar field trajectory is
well-approximated as a straight line \cite{Lehners:2006pu},
$\dot{\theta}=0$. In this case, we have $\dot{\phi}_2 = \gamma
\dot{\phi}_1,$ which implies $c_1 = \gamma c_2$. For the
colliding branes solution with empty branes, $\gamma =
\frac{1}{\sqrt{3}}\theta_H(t-t_{ref}),$ where $\theta_H$ is the
Heaviside step function (with values $\pm 1$ for $t\gtrless
t_{ref}$) and $t_{ref}$ denotes the time of the reflection, see
\cite{Lehners:2006pu} for details. The last term in equation
(\ref{eq-entropy}) is a non-local term proportional to the
difference in spatial gradients between the linear entropy
perturbation and its time derivative. This difference evolves
as $a^{-3}$ \cite{Langlois:2006vv}, so that it remains
approximately constant during the ekpyrotic phase when $a$ is
very slowly varying. This ends up being exponentially
suppressed compared to the entropy perturbation itself, which
grows by a factor of $10^{30}$ or more during this same period
\cite{Lehners:2007ac}.  After the ekpyrotic phase has ended,
the non-local term grows because $a \propto (-t)^{-1/3}$, but
this growth is negligible compared to the exponential
suppression during the ekpyrotic phase.  Hence, the non-local
term can be safely neglected. For analyzing the equation of
motion, it is convenient to use  conformal time $\tau$.
Denoting $\tau$ derivatives with primes, and introducing the
re-scaled entropy field $\delta S = a(\tau) \,\delta s,$ to
first order in perturbations, equation (\ref{eq-entropy})
reduces to ${\delta S_L}'' + \left(k^2 -\frac{a''}{a}
  + a^2 V_{ss}
 \right) \delta S_L = 0.$
The fast-roll parameter $\epsilon \equiv \dot\s^2/(2H^2)
\approx c_1^2/(2(1+\gamma^2))$ defines the equation of state of
the background scaling solution in the ekpyrotic contraction
phase. In terms of $\epsilon$ and its derivative with respect
to the number ${\cal{N}}= \ln(a_{end} H_{end}/aH)$ of e-folds
remaining before the end of the ekpyrotic phase, the solution
to the equation of motion at long wavelengths and linear order
is \be \delta S_L=f_L(k, \epsilon) \,
(-\t)^{-1+1/\epsilon-\frac{d \ln \epsilon}{2d{\cal{N}}}}. \ee
 Since $\epsilon \propto (w+1) \gg 1$ during the ekpyrotic contraction phase,
this leads to a nearly scale-invariant spectrum with index
\cite{Lehners:2007ac} $n_s = 1+ \frac{2}{\epsilon } -\frac{d
\ln \epsilon}{d{\cal{N}}}.$ As discussed in
\cite{Lehners:2007ac}, achieving a red spectral tilt consistent
with best-fit measurement ($n_s \approx 0.97$) typically
requires $\epsilon \sim {\cal{N}}^{1.5-2}$ or $c_1 \gtrsim 30.$

On large scales and at linear order, the comoving curvature
perturbation evolves according to \cite{Gordon:2000hv}
$\dot{\cal{R}}=\frac{2H}{\dot{\s}}\dot{\theta}\delta s.$ Thus, the
entropy perturbation generates a curvature perturbation when the
background trajectory bends at the reflection. Moreover, this
curvature perturbation is of the right amplitude if the potentials
turn off at a time $t_{end} \approx - 10^3 M_{Pl}^{-1},$ where
$M_{Pl}$ is the Planck mass \cite{Lehners:2007ac}.

Next we consider the non-linearities responsible for the
non-gaussianity of the perturbation spectrum. The intrinsic
non-gaussianity in the entropy perturbation is produced during
the ekpyrotic phase and can be determined from the equation of
motion for the entropy field to leading order \be \delta
S''+\left(k^2-\frac{2}{\t^2}\right)\delta S +
\frac{1}{2}aV_{sss} (\delta S)^2 = 0. \label{EomSecondOrder}\ee
The last term is approximately \be \frac{1}{2}a V_{sss} \approx
-\frac{\k_3}{2(-\tau)^2} \sqrt{\e}. \ee The solution to the
equation of motion (\ref{EomSecondOrder}), at long wavelengths
and up to second order in field perturbations, is then given by
\bea \delta S &=& \delta
S_L + \tilde{c} (\delta S_L)^2, \label{entropysecondorder} \\
\tilde{c} &=& \frac{\k_3\sqrt{\e}}{8}.\eea So $\tilde{c}$
characterizes the intrinsic non-linearity in the entropy
perturbation. As a reference, we note that for empty branes,
and with the original potential (\ref{potentialoriginal}), {\it
i.e.} with constant $c_1=-c_2/\sqrt{3}$, we have
$\k_3=-4\sqrt{2/3}$ hence $ \tilde{c} = -\sqrt{\epsilon/6} =
-c_1/4.$

The time evolution of the curvature perturbation to second
order in field perturbations and at long wavelengths is given
in FRW time by \cite{Langlois:2006vv}: \be \dot{\cal{R}} =
\frac{2H}{\dot{\s}}\dot{\theta}\delta s + \frac{H}{\dot{\s}^2}[-
(V_{ss} + 4 \dot{\theta}^2) (\delta s^{(1)})^2+\frac{V_{,\sigma}}{\dot\s}\d s \dot{\d s}].
\label{zetadotquadratic} \ee At linear order a bending
($\dot{\theta}\neq 0$) of the background trajectory combined
with a non-zero entropy perturbation source the curvature
perturbation on large scales and result in a linear, gaussian
curvature perturbation \be {\cal{R}}_L= \int
\frac{2H}{\dot{\s}}\dot{\theta}\delta s^{(1)}.
\label{curvaturelinear} \ee During this process of conversion,
the entropy perturbation evolves according to Eq.
(\ref{eq-entropy}) in a rather complicated way, which is why
the above integral and the following calculation have to be
performed numerically. Qualitatively, a sharp reflection leads
to a drastic diminution in the entropy perturbation and a much
smaller value of the above integral, while a smooth and more
gradual reflection results in an enhanced efficiency of
conversion.

The leading non-gaussianity is generated when modes are outside
the horizon, so the non-gaussianity is of the local,
wavelength-independent type
\cite{Komatsu:2000vy,Maldacena:2002vr,Sefusatti:2007ih}. The
non-linearity parameter $f_{NL}$ is then defined by the
relations \cite{{Komatsu:2000vy}} $\Phi_H = \Phi_L + f_{NL}
\Phi_L^2,$ where $\Phi_H$ is Bardeen's space-space metric
perturbation \cite{Bardeen:1980kt} and $\Phi_L$ is its value to
linear order. We adopt the convention of
Ref.~\cite{Langlois:2006vv} where the  comoving curvature
perturbation is equal but opposite in sign to Bardeen's $\zeta$
variable on large scales, so that during matter domination we
have ${\cal{R}} = -\frac{5}{3}\Phi_H$, and thus
${\cal{R}}={\cal{R}}_L-\frac{3}{5}f_{NL}{\cal{R}}_L^{2}.$ (Note
that this definition of $f_{NL}$ agrees with Komatsu and
Spergel \cite{Komatsu:2000vy} and WMAP \cite{Spergel:2006hy}
but is opposite in sign
 to Maldacena \cite{Maldacena:2002vr}.)

At the quadratic level, Eq. (\ref{zetadotquadratic}) then
implies that there are three distinct ways in which the
curvature perturbation can acquire non-linear contributions:

(1) Before the reflection, the terms proportional to $V_{ss}$ and $V_{,\s}$
create non-linearities in $\cal{R}$ during the entire
ekpyrotic phase in which the entropy perturbation is generated;
during this phase, the background trajectory is a straight
line, so Eq.~(\ref{zetadotquadratic}) reduces to $\dot{\cal{R}}
= \frac{H}{\dot{\s}^2} (-V_{ss} (\delta s)^2+\frac{V_{,\s}}{\dot\s}\d s\dot{\d s}),$ which to leading
order reads $\dot{\cal{R}} = \frac{1}{2}f_L^{2}(-t)^{-3}.$ The integrated
contribution thus amounts to \be {\cal{R}}_{integrated} =
-\frac{1}{4} (\delta
s_{end}^{(1)})^2.
\label{entropyintegrated}\ee This term leads to \be
f_{NL}^{integrated}=+\frac{5}{12 {\cal{R}}_L^2}(\delta
s_{end}^{(1)})^2, \ee and is typically small, of order $O(10)$
or less.

(2) At the reflection, any intrinsic non-linearity in $\delta
s$ will be converted into a non-linearity in $\cal{R}$ due to
the first term in (\ref{zetadotquadratic}); thus we get \be
f_{NL}^{intrinsic}=-\frac{5}{3{\cal{R}}_L^2}\int
\frac{2H}{\dot{\s}}\dot{\theta}\delta s^{(2)}. \ee This term is
positive and increases with increasing $\tilde{c},$ although
generally not quite as fast as linearly because of the non-linear
evolution of $\delta s^{(2)}.$

(3) Also at the reflection, the terms involving $\delta s^{(1)}$
lead to non-linearities in $\cal{R}$ according to \bea
f_{NL}^{reflection}=\frac{5}{3{\cal{R}}_L^2}\int_{ref}
\frac{H}{\dot{\s}^2} &&[(V_{ss} + 4 \dot{\theta}^2) (\delta
s^{(1)})^2 \nn \\ && -\frac{V_{,\s}}{\dot\s}\d s\dot{\d s}]. \label{fNLreflection} \eea In order to close the equations for the
purposes of this study, we
are modeling the reflection in terms of an effective potential $V$
that acts when $\phi_{2}$ gets small and causes it to reflect; we
find that the results do not depend sensitively on the form of $V$
except for how gradual the reflection is. Even
disregarding the ${{\cal{R}}_L^{-2}}$ prefactor, this term
generically increases in magnitude as the reflection is made
sharper and it is always negative in sign.

\begin{figure}[t]
\begin{center}
\includegraphics[width=0.45\textwidth]{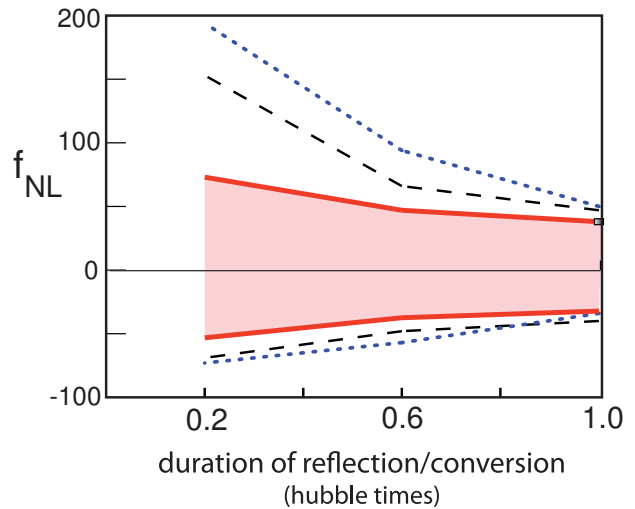}
\caption{\label{Figure2}
The predicted value of the non-gaussianity parameter $f_{NL}$ as a
function of the duration of the reflection.  The shaded range
represents results for the simplest ekpyrotic and reflection
potentials allowing $\kappa_3$ to range from $+5$ (top) to $-5$
(bottom).  The upper and lower boundaries can be extended (dashed and
dotted curves) by carefully tuning the reflection potentials, but the
change is not significant for typical durations ${\cal O} (1)$.}
 \end{center}
\end{figure}

The total $f_{NL}$ is the sum of all the above contributions.
In Fig.~\ref{Figure2}, we have shown the results for a typical value $\e=36$
(corresponding to {\it e.g.} $c_1=c_2=12$) and for the range
$-5 \leqslant \k_3 \leqslant 5$ with the lower curves
corresponding to $\k_3=-5$ and the higher curves to $\k_3=+5.$
In order to see how robust the results are, we experiment with
different potential forms to model the reflection of $\phi_2$.
The results shown here are for the potential forms
$V_1(\phi_2)= v (\phi_2^{-2} +r \phi_2^{-6})$ and $V_2(\phi_2)
= v[({\rm sinh}\, \phi_2)^{-2} + r ({\rm sinh}\,
\phi_2)^{-4}]$, where we have varied the coefficient $v$ over
several orders of magnitude and taken $r=0,1$ in each case.
Each potential is represented in Fig.~\ref{Figure2} by a
different curve, with the $r=0$ examples corresponding to the
curves bounding the shaded area.  We will not give further
details here because, as we will show, they are not important.

We find that the results for $f_{NL}$ fall into two distinct
regimes depending mainly on the duration of the conversion
process. We measure the duration by specifying the number
of e-folds (`Hubble times') by which the scale factor $a$
shrinks during the process of conversion; it is useful to
distinguish the case of ``rapid conversion,'' where the conversion lasts at most about $0.2$ Hubble times,  from the case of ``gradual
conversion,'' where the conversion lasts on the order of $1$ Hubble time
(The details of the reflection affect the direction of the
trajectory afterwards, which  has to be taken into account when
incorporating this conversion mechanism within the full
cosmological model; but this is beyond the scope of this
paper.)

In the case of rapid conversion, the conversion is inefficient so that the
leading order
curvature contribution ${{\cal{R}}_L}$ is smaller by a factor of
${\cal{O}}(10)$ or more compared to the case of gradual conversions.
Reducing
${{\cal{R}}_L}$ by a factor of ten automatically amplifies  all
the contributions to $f_{NL}$ by a factor of ${{\cal{O}}}(100),$
although the summing up of all the contributing terms cancels out some of the amplification.
The net result for $f_{NL}$ turns out to be of ${\cal{O}}(100)$ or more,
as shown in Fig.~\ref{Figure2} for the case when the duration of the conversion
is only $0.2$ Hubble times.  The figure
also shows that the result is sensitive to the form of the potential and the
potential parameters ($v$ and $r$) used to model the reflection.  However, much of the
entire range lies outside the observationally acceptable range, and so can be
dropped from further consideration. Here we are chiefly interested in making general predictions, and not in the
accidental cancellation of large numbers, which would clearly be fine-tuned and
unattractive.

The second regime is where the reflection and conversion
process are gradual. In these cases, the conversion is
efficient and all contributions to $f_{NL}$ tend to be reduced.
Roughly, for $\k_3$ ranging from $-5$ to $5$ and for typical values of $\e \sim 50,$  we find that the predictions converge to a
range  \be \label{result} -60 \lesssim f_{NL} \lesssim +80,
\ee with very weak dependence on the parameters of the
reflection potential. If $\e$ is pushed to higher values, {\it e.g.} up to $\e =200,$ this range expands to $-120 \lesssim f_{NL} \lesssim +160.$ Analytic
estimates of the above calculations confirm that the results
are rather insensitive to the specific form of the potential
causing the reflection, but depend most strongly on the
duration of conversion as well as roughly linearly on
$\k_3{\sqrt{\e}}$ \cite{LehnersSteinhardtInPrep}.

Several comments are in order.  First, our results show that, in the ekpyrotic
model,
it is very unnatural to get
a value of $|f_{NL}| \lesssim 1$, the value obtained for slow-roll
inflation \cite{Maldacena:2002vr}. The contributions to the local $f_{NL}$
are generally greater than ten so that small values can only be
achieved by accidental cancellation of large terms.

Conversely,
obtaining large local $f_{NL}$ is completely non-generic for
inflation, requiring the finely-tuned addition of extra fields and interactions.
When the fields and interactions are introduced, there is no predictive
convergence to a bounded range for $f_{NL}$.  The result can be anything from
negligible $f_{NL}$ to values of either sign and extending orders of magnitude
beyond current bounds.
By contrast, in
the ekpyrotic model, conversion of scalar fluctuations to
curvature perturbations is absolutely required, and, at least for one proven
conversion mechanism, generically generates a measurable $f_{NL}$.
We have found that a
rapid conversion, whether in an ekpyrotic or kinetic energy
dominated phase, generically leads to a large $|f_{NL}| = {\cal O}
(100)$ or greater that is already ruled out observationally. Furthermore, if
the conversion occurs in the ekyprotic phase, it is difficult to obtain
simultaneously an observationally acceptable $f_{NL}$ and an observationally
acceptable spectral tilt.  On
the other hand, a wide range of ekpyrotic/cyclic models
satisfying current constraints on the spectral tilt generate a
measurable non-gaussianity consistent with current observations if
the conversion is gradual and occurs after the ekpyrotic phase is
completed.

\noindent {\it Acknowledgements} We would like to thank T.
Battefeld, D. Baumann, J. Khoury, E. Komatsu, N. Turok, F. Vernizzi for useful
discussions and particularly K. Koyama and D. Wands for remarks
that pointed us to the differences between rapid and gradual
conversion, and S. Renaux-Petel for pointing out a sign error in an earlier version of this paper. This  work is supported in part by the US Department
of Energy grant DE-FG02-91ER40671.

\end{document}